\documentclass[conference]{IEEEtran}

\makeatletter
\def\bstctlcite{\@ifnextchar[{\@bstctlcite}{\@bstctlcite[@auxout]}}
\def\@bstctlcite[#1]#2{\@bsphack
  \@for\@citeb:=#2\do{%
    \edef\@citeb{\expandafter\@firstofone\@citeb}%
    \if@filesw\immediate\write\csname #1\endcsname{\string\citation{\@citeb}}\fi}%
  \@esphack}
\makeatother
\usepackage{multicol}
\usepackage{multirow}
\usepackage{graphicx}
\usepackage{amsmath}
\usepackage{amsfonts}
\usepackage{amssymb}
\usepackage{amsthm}
\usepackage{tikz}
\usepackage{bm}
\usepackage{makecell}
\usepackage{caption}
\newtheorem{lemma}{Lemma}

\newtheorem{theorem}{Theorem}
\newtheorem{corollary}{Corollary}
\newtheorem{construction}{Construction}
\newtheorem{remark}{Remark}

\begin{document}
\title{Rack-Aware Regenerating Codes with Fewer\\ Helper Racks}

\author{\IEEEauthorblockN{Zhifang Zhang, Liyang Zhou}
\IEEEauthorblockA{\fontsize{9.8}{12}\selectfont KLMM, Academy of Mathematics and Systems Science, Chinese Academy of Sciences, Beijing 100190, China\\
School of Mathematical Sciences, University of Chinese Academy of Sciences, Beijing 100049, China\\
Emails: zfz@amss.ac.cn,~~zhouliyang17@mails.ucas.ac.cn}
}
\maketitle

\thispagestyle{empty}
\begin{abstract}
We consider the rack-aware storage system where $n$ nodes are organized in \(\bar{n}\) racks each containing \(u\) nodes, and any $k$ nodes can retrieve the stored file. Moreover, any single node erasure can be recovered by downloading data from $\bar{d}$ helper racks as well as the remaining $u\!-\!1$ nodes in the same rack. Previous work mostly focuses on minimizing the cross-rack repair bandwidth under the condition $\bar{d}\geq \bar{k}$, where $\bar{k}=\lfloor\frac{k}{u}\rfloor$. However, $\bar{d}\geq \bar{k}$ is not an intrinsic condition for the rack-aware storage model. In this paper, we establish a tradeoff between the storage overhead and cross-rack repair bandwidth for the particularly interesting case $\bar{d}\!<\!\bar{k}$. Furthermore, we present explicit constructions of codes with parameters lying on the tradeoff curve respectively at the minimum storage point and minimum bandwidth point. The codes are  scalar or have sub-packetization $\bar{d}$, and operate over finite fields of size comparable to $n$. Regarding $\bar{d}$ as the repair degree, these codes combine the advantage of regenerating codes in minimizing the repair bandwidth and that of locally repairable codes in reducing the repair degree. Moreover, they also abandon the restriction of MBR codes having storage overhead no less than $2\times$ and that of high-rate MSR codes having exponential sub-packetization level.

\end{abstract}


\section{Introduction}\label{sec0}
\IEEEPARstart{D}{eployment} of erasure codes in distributed storage systems can ensure fault-tolerant storage with low redundancy, however, the node repair problem becomes an important issue. Two metrics of the repair efficiency are {\it repair bandwidth} (i.e., the amount of data transmitted in the repair process) and {\it repair degree} (i.e., the number of nodes connected in the repair process). In order to optimize these metrics, regenerating codes (RC) \cite{Dimakis2011} and locally repairable codes (LRC) \cite{LRC} are respectively proposed. Let us first recall the model for RC. Suppose a data file is stored across $n$ nodes such that any $k$ nodes can together retrieve the data file. Any single node erasure can be recovered by downloading data from any $d$ surviving nodes (i.e., helper nodes). Due to the minimality of $k$ \footnote{I.e., there exist $k-1$ nodes that cannot retrieve the data file.}, this model naturally implies $d\geq k$. By contrast to the naive repair approach which recovers the whole data file to repair a node failure, RC reduces the repair bandwidth at the cost of connecting to more helper nodes. Nevertheless, it achieves the optimal tradeoff between the storage overhead and repair bandwidth. A lot of work devote to studying RCs, especially the ones with the minimum storage (MSR codes) and with the minimum bandwidth (MBR codes) \cite{Kumar2011,Ye2016,Balaji2018Overview}. Alternatively, in some scenarios the repair degree becomes a more sensitive metric and the $[n,k]$ LRCs where each node erasure can be recovered by connecting to $r\!\ll\! k$ helper nodes are more desirable \cite{XORing2013,LRCFamily2014}. In pursuit of small repair degree, LRC sacrifices the fault-tolerance level and also leaves out the repair bandwidth. In \cite{Govinda14LocalReg} the authors proposed the codes with local regeneration which combine RC and LRC by designing each local code as an MSR or MBR code. However, since MSR/MBR codes with small parameters only have low rate, the resultant code cannot possess practical storage overhead.

Surprisingly, we find optimization of the two metrics naturally merge in the rack-aware storage system. Specifically, suppose $n$ nodes are organized in \(\bar{n}\) racks each containing \(u\) nodes. Any file stored in the system can be retrieved by connecting to any $k$ out of the $n$ nodes. That is, the fault tolerance of the system is $n\!-\!k$. Moreover, any single node erasure can be recovered by downloading data from $\bar{d}$ helper racks as well as the remaining $u-1$ nodes in the same rack. More importantly, the cost of cross-rack communication is far more expensive than that of the intra-rack communication. Therefore, we calculate the repair bandwidth by only counting the cross-rack repair bandwidth and view $\bar{d}$ as the repair degree. Our goal is to minimize the repair bandwidth when $\bar{d}$ is small.

The problem of minimizing the cross-rack repair bandwidth was proposed in \cite{Hu} for hierarchical data centers.
Suppose $k\!=\!\bar{k}u\!+\!u_0,~0\!\leq\!u_0\!<\!u$. Hou et al. \cite{Hou} firstly derived
the tradeoff between the storage overhead and repair bandwidth under the condition $\bar{d}\geq\bar{k}$. In particular, the two extreme points on the tradeoff curve respectively correspond to the minimum storage rack-aware regenerating (MSRR) codes and minimum bandwidth rack-aware regenerating (MBRR) codes. Then in \cite{Chen} explicit constructions of MSRR codes were developed. Recently, Hou et al. \cite{Hou2020} further reduced the sub-packetization level for MSRR codes. Besides, some work studied the clustered storage system which is almost the same with the rack-aware system, except that the repair bandwidth measurement and data reconstruction requirement vary in different papers \cite{Prakash2018,Sohn2019}.

\subsection{$\bar{d}\geq \bar{k}$ is not intrinsic for rack-aware storage}
Although the work of \cite{Hou,Chen,Hou2020,Sohn2019} all restricted to the case $\bar{d}\geq \bar{k}$, we note $\bar{d}\!\geq\! \bar{k}$ is not an intrinsic condition for the rack-aware storage model. Because the surviving $u\!-\!1$ nodes located in the same rack with the failed node must participate in the repair, $\bar{d}<\bar{k}$ will not contradict the minimality of $k$. Therefore, it is possible to have the repair degree $\bar{d}\ll\bar{k}$ in the rack-aware storage system. Particularly when $\bar{d}=0$, this rack-aware storage model coincides with that of LRCs where the $u$ nodes within each rack form a local repair group.

Paper \cite{Prakash2018} considered minimizing the inter-cluster (i.e., cross-rack) repair bandwidth in the case $\bar{d}<\bar{k}$. However, in their clustered storage system the data reconstruction is realized by connecting to any $\bar{k}$ clusters (i.e., racks) rather than any $k$ nodes. As a result, their codes only have fault tolerance $\bar{n}-\bar{k}$ which is rather weak when the storage overhead is simultaneously kept in a low level. By contrast, our codes have fault tolerance $n-k\approx (\bar{n}-\bar{k})u$ and storage overhead near to $\bar{n}/\bar{k}$ for properly chosen parameters.

\subsection{Our Contributions}
With respect to the rack-aware storage system introduced before, we firstly derive a cut-set bound on the maximum size of the file that can be stored in the system given constraints on the storage per node and cross-rack bandwidth. The cut-set bound applies to all parameters, i.e., $n\!=\!\bar{n}u, ~k\!=\!\bar{k}u+u_{0},~0\!\leq \!u_{0}\!<\!u$ and $0\!\leq\!\bar{d}\!<\!\bar{n}$. When $\bar{d}\geq\bar{k}$ it coincides with the one obtained in \cite{Hou}.

The cut-set bound induces a tradeoff between the storage overhead and repair bandwidth. Furthermore, we characterize parameters of the two extreme points on the tradeoff curve that possess the minimum storage and minimum bandwidth respectively. For simplicity,  we also term the codes at the extreme points  MSRR and MBRR codes respectively as in \cite{Hou}, although here we focus on the case $0\leq \bar{d}<\bar{k}$.

Then we present explicit constructions of MSRR codes for $0\leq \bar{d}<\bar{k}$ and MBRR codes for $0<\bar{d}<\bar{k}$. Note when $\bar{d}\!=\!0$ there is no cross-rack communication in the repair process, so we consider no MBRR code for $\bar{d}=0$. Both of the MSRR and MBRR codes are built over a finite field $F$ satisfying $u\mid (|F|-1)$ and $|F|>n$. Moreover, the MSRR code is scalar over $F$ and MBRR code is an array code over $F$ with sub-packetization $\bar{d}$. Under properly chosen parameters, our codes can possess acceptable fault tolerance and practical storage overhead in addition to the optimal repair bandwidth.

The remaining of the paper is organized as follows. Section II derives the cut-set bound and the parameters for MSRR and MBRR codes. Section III and IV respectively present the explicit constructions of MSRR and MBRR codes. Section V concludes the paper.

\section{The cut-set bound}\label{sec1}
First we introduce some notations. For nonnegative integers $m<n$, let $[n]=\{1,...,n\}$ and $[m,n]=\{m,m+1,...,n\}$. Throughout the paper we label the racks from $0$ to $\bar{n}-1$ and the nodes within each rack from $0$ to $u\!-\!1$. Moreover, we represent each node by a pair $(e,g)\in[0,\bar{n}\!-\!1]\times [0,u\!-\!1]$ where $e$ is the rack index and $g$ is the node index within the rack.
Then we describe an information flow graph to illustrate the node repair and file reconstruction in rack-aware storage system.
\begin{itemize}
\item The data file consisting of $B$ symbols flows from the source vertex S to each data collector C which connects to $k$ nodes. Each node $(e,g)$ is split into two nodes $X^{\rm in}_{\scriptscriptstyle{(e,g)}}$ and $X^{\rm out}_{\scriptscriptstyle{(e,g)}}$ with a directed edge of capacity $\alpha$.

\item When a node $X_{\scriptscriptstyle{(e,g)}}$ fails, the replacement node $X'_{\scriptscriptstyle{(e,g)}}$ connects to $\bar{d}$ helper racks and the remaining $u\!-\!1$ nodes in rack $e$.  The surviving nodes in rack $e$ are copied to the replacement rack with edges of capacity $\infty$ from the original node to the copy.

\item Within each helper rack, there exist directed edges of capacity $\infty$ from the other nodes to the connected node indicating the free intra-rack communication.

\end{itemize}

\begin{figure}[ht]
\begin{center}
\includegraphics[width=0.9\columnwidth]{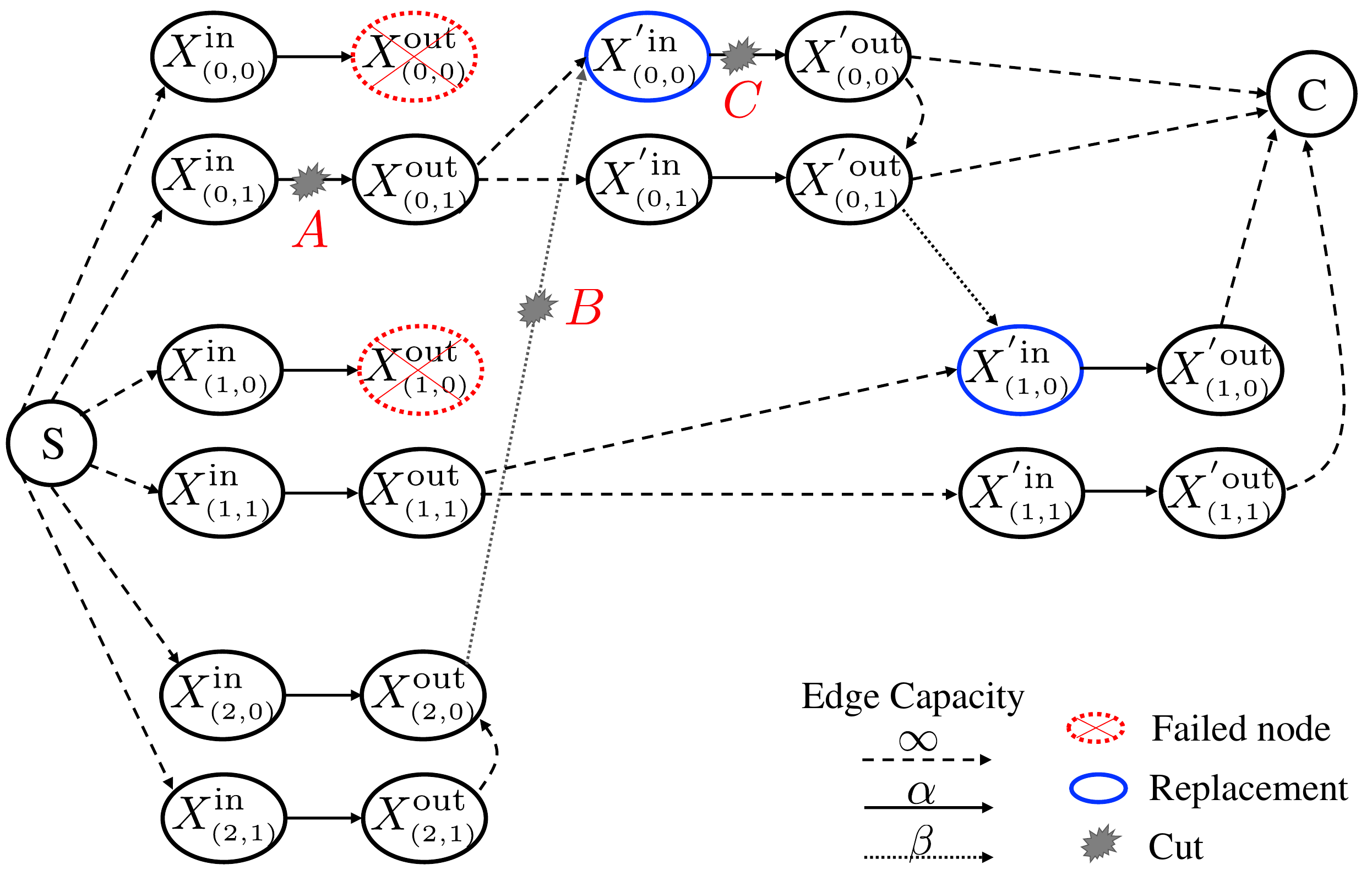}
\end{center}
\caption{\scriptsize An example of the information flow graph used to prove the cut-set bound, where $(n,k,u,\bar{d})=(6,4,2,1)$.}\label{fg1}
\end{figure}

\begin{theorem}(Cut-set bound) For a rack-aware storage system with parameters $(n,k,u,\bar{d};\alpha,\beta,B)$, let $B^*$ denote the maximum possible file size, then
\begin{equation}\label{cutsetbound}
B\leq B^*\!=\! (k-\bar{k})\alpha+\textstyle{\sum_{i=1}^{\min\{\bar{k},\bar{d}\}}}\min\{(\bar{d}-i+1)\beta,\alpha\}\;.	
\end{equation}	
\end{theorem}
\begin{proof}
By network coding for the multicast problem \cite{Yeung08}, $B^*\!=\!\min_{\scriptscriptstyle{\rm C}}$MinCut(S,C). Therefore, it is sufficient to figure out the data collector C having the minimum MinCut(S,C). Suppose C connects to $\bar{l}$ racks and $v$ single nodes. Note some of the $v$ nodes may locate in the same rack, but not occupy the whole rack. Obviously, $\bar{l}u+v=k$ and $\bar{l}\leq \bar{k}$.

For the first rack connected by C, if no failure happens then it contributes $u\alpha$ to MinCut(S,C), else it contributes $(u-1)\alpha+\min\{\bar{d}\beta,\alpha\}$. Take the information flow graph in Fig. \ref{fg1} for example, to cut rack $0$ from S there are two possible MinCut, i.e., $\{A,B\}$ or $\{A,C\}$. The former has cut value $\alpha+\beta$ and the latter has $2\alpha$. Note this observation holds even for the case $\bar{d}=0$ where $\beta=0$.
For the second rack, it contributes $(u-1)\alpha+\min\{(\bar{d}-1)\beta,\alpha\}$ because the first rack can serve as a helper rack for the node failure in the second rack and this helper rack has already been cut. Following this analysis, the $\bar{l}$ racks contribute $\bar{l}(u-1)\alpha+\sum_{i=1}^{\min\{\bar{l},\bar{d}\}}\min\{(\bar{d}-i+1)\beta,\alpha\}$.

Additionally, suppose C connects to $v'$ nodes in the $(\bar{l}+1)$-th rack where $v'<u$. If a failure happens among the $v'$ nodes, they contribute $(u\!-\!1)\alpha\!+\!\min\{(\bar{d}-x)\beta,\alpha\}$ to MinCut(S,C) where $x=\min\{\bar{l},\bar{d}\}$, else they contribute $v'\alpha$ which is less than the former. Therefore, the $v$ single nodes connected by C contribute $v\alpha$ in order to achieve ${\min_{\scriptscriptstyle{\rm C}}}$MinCut(S,C).

As a result,
{%\renewcommand*{\arraystretch}{1.5}
	\setlength{\abovedisplayskip}{6pt}
	\setlength{\belowdisplayskip}{6pt}
\begin{eqnarray}
	&&\mbox{MinCut(S,C)}\notag\\&=&(\bar{l}(u-1)+v)\alpha+\textstyle{\sum_{i=1}^{\min\{\bar{l},\bar{d}\}}\min\{(\bar{d}-i+1)\beta,\alpha\}}\notag\\&=&(k-\bar{l})\alpha+\textstyle{\sum_{i=1}^{\min\{\bar{l},\bar{d}\}}\min\{(\bar{d}-i+1)\beta,\alpha\}}\label{eq22}
\end{eqnarray}}

\noindent One can see the value of (\ref{eq22}) decreases as $\bar{l}$ grows. Therefore when $\bar{l}=\bar{k}$ it reaches the minimum MinCut among all data collectors. Thus the theorem is proved for all $0\leq\bar{d}<\bar{n}$.
\end{proof}

Particularly when $\bar{d}\!\geq\!\bar{k}$ the cut-set bound (\ref{cutsetbound}) coincides with the one obtained in \cite{Hou}. When $\bar{d}\!<\!\bar{k}$, from (\ref{cutsetbound}) it has
\begin{eqnarray}
B&\leq&  (k-\bar{k})\alpha+\textstyle{\sum_{i=1}^{\bar{d}}\min\{(\bar{d}-i+1)\beta,\alpha\}}\notag\\
&\leq& (k-\bar{k}+\bar{d})\alpha\;.\label{eq3}
\end{eqnarray}
Meanwhile, the total size of the downloaded data for repairing a  failed node $(e^*,g^*)$ is $(u-1)\alpha+\bar{d}\beta$. From the downloaded data it can actually recover all $u$ nodes in rack $e^*$. Thus it follows $(u-1)\alpha+\bar{d}\beta\geq u\alpha$, i.e., $\bar{d}\beta\geq \alpha$. Combining with (\ref{cutsetbound}) and (\ref{eq3}) we characterize the two extreme points on the $\alpha$-$\beta$ tradeoff curve in the following corollary.

\begin{corollary}
Suppose $0\!<\!\bar{d}\!<\!\bar{k}$. The MSRR and MBRR codes have the following parameters:
\begin{equation}\label{msrr}
\alpha_{\scriptscriptstyle{\rm MSRR}}=\beta_{\scriptscriptstyle{\rm MSRR}}=B/(k-\bar{k}+\bar{d})\;,
\end{equation}
\begin{equation}\label{mbrr}
\alpha_{\scriptscriptstyle{\rm MBRR}}=\bar{d}\beta_{\scriptscriptstyle{\rm MBRR}}=B/(k-\bar{k}+\frac{\bar{d}+1}{2})\;.
\end{equation}
Particularly when $\bar{d}=0$, it has $\beta=0$ and we only consider MSRR codes which have $\alpha_{\scriptscriptstyle{\rm MSRR}}=B/(k-\bar{k})$.
\end{corollary}

\section{Construction of MSRR codes for $0\leq\bar{d}<\bar{k}$}
Assume $\beta=1$ (for $\bar{d}>0$), then according to \eqref{msrr} the MSRR codes have $\alpha=1$ and $B=k-\bar{k}+\bar{d}$. Actually, the MSRR code we construct is an $[n,B]$ linear code $\mathcal{C}$ over a finite field $F$ satisfying $u|(|F|-1)$ and $|F|>n$.
First introduce some notations.
\begin{itemize}
\item Denote $T_1\!=\!\{iu\!\mid\! i\!\in\![0,\bar{n}\!-\!\bar{d}\!-\!1]\}$ and $T\!=\![0,n\!-\!k\!-\!1]\cup T_1$. It is easy to check $|T|=n-k+\bar{k}-\bar{d}=n-B$.
\item Let $\xi$ be a primitive element of $F$ and $\eta$ be an element of $F$ with multiplicative order $u$.
\item Denote $\lambda_{(e,g)}\!=\!\xi^e\eta^g$ for $e\!\in\![0,\bar{n}\!-\!1],g\!\in\![0,u\!-\!1]$. It can be seen $\lambda_{(e,g)}\!\neq\!\lambda_{(e',g')}$ for $(e,g)\!\neq\! (e',g')$, because $(\xi^{e-e'})^u\neq 1$ for $e\!\neq\! e'\in[0,\bar{n}-1]$ while $(\eta^{g'\!-g})^u=1$ for all $g,g'\in[0,u-1]$.
\end{itemize}

\begin{construction}\label{cons1}
Represent each codeword in $\mathcal{C}$ by a vector ${\bm c}\!=\!({\bm c}_0,...,{\bm c}_{\bar{n}-1})\!\in\! (F^u)^{\bar{n}}$, where ${\bm c}_e\!=\!(c_{(e,0)},...,c_{(e,u-1)})\!\in\! F^u$ and $c_{(e,g)}$ is stored in node $(e,g)$ for $e\in[0,\bar{n}-1]$ and $g\in[0,u-1]$. Then the code $\mathcal{C}$
is defined as follows
\begin{equation}\label{PCE}
\mathcal{C}=\{{\bm c}\in F^n\mid \sum_{e=0}^{\bar{n}-1}\sum_{g=0}^{u-1}\lambda^t_{(e,g)}c_{(e,g)}=0,\ \forall t\!\in\! T\}\;.
\end{equation}
\end{construction}
For simplicity, consider the $n\times n$ Vandermonde matrix $(\lambda^t_{(e,g)})_{t,(e,g)}$ with row index $t\in[0,n\!-\!1]$ and column index $(e,g)\in[0,\bar{n}\!-\!1]\!\times\![0,u\!-\!1]$, among which the rows indexed by the set $T$ actually forms a parity check matrix of $\mathcal{C}$. Thus $\mathcal{C}$ is evidently an $[n,B]$ linear code. Furthermore, since $[0,n-k-1]\subseteq T$, $\mathcal{C}$ is a $B$-dimensional subcode of an $[n,k]$ generalized Reed-Solomon (GRS) code. As a result, any $k$ nodes can recover the original data file. Next, we prove the repair property of $\mathcal{C}$.

\begin{theorem}
The code $\mathcal{C}$ in Construction \ref{cons1} satisfies the optimal repair property, i.e., for any node $(e^*,g^*)$ and any $\mathcal{H}\!\subseteq\! [0,\bar{n}\!-\!1]\!-\!\{e^*\}$ with $|\mathcal{H}|\!=\!\bar{d}$, the symbol $c_{(e^*,g^*)}$ can be recovered from
$$\{c_{(e^*,g)}\mid g\in[0,u-1],g\neq g^*\}\cup\{s_e\mid e\in\mathcal{H}\}$$
where $s_e\!\in F$ is computed from $\{c_{(e,g)}\mid g\in[0,u-1]\}$.
\end{theorem}
\begin{proof}
For simplicity, we denote
\begin{equation}\label{cs}
\tilde{c}_e=\textstyle{\sum_{g=0}^{u-1}c_{(e,g)}}\in F, ~\forall e\in[0,\bar{n}-1]\;.
\end{equation}
Obviously, $c_{(e^*,g^*)}$ can be easily computed from $\tilde{c}_{e^*}$ and the intra-rack transmission $\{c_{(e^*,g)}\!\mid\! g\!\in\![0, u-1], g\!\neq\! g^*\}$. Therefore, it is sufficient to prove that
$\tilde{c}_{e^*}$ can be recovered from $\{s_e\mid e\in\mathcal{H}\}$.

We select a system of the parity check equations with respect to the values of $t\in T_1$, i.e.,
\begin{equation}\label{R1}
\sum_{e=0}^{\bar{n}-1}\sum_{g=0}^{u-1}\lambda_{(e,g)}^{iu} c_{(e,g)}=0,~~ i\!\in\![0,\bar{n}\!-\bar{d}\!-\!1]\;.
\end{equation}
Furthermore, because $\lambda_{(e,g)}=\xi^e\eta^g$ and $\eta$ has multiplicative order $u$, it has $\lambda_{(e,g)}^{iu}=(\xi^{eu})^i$. Then using the notation defined in \eqref{cs}, the linear system \eqref{R1} becomes
\begin{equation}\label{R2}
(\xi^{e^*u})^i\tilde{c}_{e^*}\!+\!\textstyle{\sum_{e\neq e^*}(\xi^{eu})^i\tilde{c}_{e}}\!=\!0,~ i\!\in\![0,\bar{n}\!-\bar{d}\!-\!1]\;,
\end{equation}
which indicates $(\tilde{c}_0,\tilde{c}_1,...,\tilde{c}_{\bar{n}-1})$ is a codeword of an $[\bar{n},\bar{d}]$ GRS code. Therefore, the symbol $\tilde{c}_{e^*}$ can be recovered from the $\bar{d}$	symbols $\{\tilde{c}_{e}\mid e\in\mathcal{H}\}$ which are downloaded from the helper racks, i.e., $s_e=\tilde{c}_{e}$.
\end{proof}

\begin{remark}\label{re1}
In \cite{LRCFamily2014} the LRC that attains the Singleton-like bound with equality is called an optimal LRC. Next we state when $\bar{d}=0$ Construction \ref{cons1} actually defines an $[n,k-\bar{k}]$ optimal LRC with all symbol locality $u-1$. First, the linear system (\ref{R2}) implies $\tilde{c}_e=0$ for all $e\in[0,\bar{n}-1]$ in the case $\bar{d}=0$. Then by the definition of $\tilde{c}_e$ in (\ref{cs}) each node can be recovered by the remaining $u-1$ nodes in the same rack, i.e.,  has repair locality $u-1$. Next we prove achievability of the Singleton-like bound \cite{LRC}. Let $d_{\rm min}$ be the minimum distance of $\mathcal{C}$. There are two cases:
\begin{enumerate}
	\item $u\nmid k$. From the Singleton-like bound it has $d_{\rm min}\leq n-(k-\bar{k})+2-\lceil\frac{k-\bar{k}}{u-1}\rceil=n-k+1$. On the other hand, since any $k$ nodes can recover the data file, it follows $d_{\rm min}\geq n-k+1$. Thus the Singleton-like bound is met with equality.
	\item $u\mid k$.  For $x\in\mathbb{N}$ define $D(x)=x-\bar{x}$ where $\bar{x}=\lfloor\frac{x}{u}\rfloor$. One can check $D(\bar{k}u)=D(\bar{k}u-1)$. Thus in the case $k=\bar{k}u$ we actually build an $(n,k-1,\bar{d}=0)$ MSRR code which turns out to be an $[n,k-\bar{k}]$ LRC with locality $u-1$. From the Singleton-like bound,  $d_{\rm min}\leq n-(k-\bar{k})+2-\lceil\frac{k-\bar{k}}{u-1}\rceil=n-k+2$. On the other hand, since any $k-1$ nodes can recover the data file, it has $d_{\rm min}\geq n-k+2$. Thus resultant $[n,k-\bar{k}]$ linear code is an optimal LRC.
\end{enumerate}
That is, in \cite{LRCFamily2014} the authors construct optimal LRCs from the generator matrices, while our Construction \ref{cons1} provides another construction of optimal LRCs from the parity check matrices.
	
\end{remark}

\section{Construction of MBRR codes for $0<\bar{d}<\bar{k}$}
Similarly, assume $\beta=1$, then by \eqref{mbrr} the MBRR codes have
$\alpha=\bar{d}$ and $B=(k\!-\!\bar{k})\bar{d}+\frac{\bar{d}(\bar{d}+1)}{2}$.
The MBRR code we construct is a linear array code over $F$ with sub-packetization $\bar{d}$, where $F$ is a finite field satisfying $u\mid (|F|\!-\!1)$ and $|F|>n$.

First, denote $J_1=\{tu+u-1: t\in[0,\bar{k}-1]\}$ and $J_2=[0,k-1]-J_1$. Obviously, $|J_1|=\bar{k}$ and $|J_2|=k-\bar{k}$.
Then we write the $B$ symbols of the data file into a $\bar{d}\times k$ matrix $M=(m_{i,j})_{i\in[0,\bar{d}-1],j\in [0,k-1]}$. Let $M_1$ and $M_2$ denote the matrices of $M$ restricted to the columns indexed by $J_1$ and $J_2$ respectively. Moreover, $M_1$ has the following form
{\setlength\abovedisplayskip{6pt}
	\setlength\belowdisplayskip{6pt}
\begin{equation}\label{ms2}
M_1=\begin{pmatrix}
S&\bm 0\\	
\end{pmatrix},	
\end{equation}}
where $S$ is a $\bar{d}\times\bar{d}$ symmetric matrix with the the upper-triangular half filled up with $\frac{\bar{d}(\bar{d}+1)}{2}$ symbols of the data file, $\bm 0$ is a $\bar{d}\times(\bar{k}-\bar{d})$ all zero matrix.  $M_2$ is a $\bar{d}\times(k-\bar{k})$ matrix containing the remaining $(k-\bar{k})\bar{d}$ symbols of the data file.

Next, for $i\in[0,\bar{d}-1]$ define polynomials
{\setlength\abovedisplayskip{6pt}
	\setlength\belowdisplayskip{6pt}
 \begin{equation}\label{polys}f_i(x)=\textstyle{\sum_{j=0}^{k-1}m_{i,j}x^j}\;.\end{equation}}
Let $\xi,\eta,\lambda_{(e,g)}$ be the same as in Section III. Then each codeword of the MBRR code is composed of values of the $\bar{d}$ polynomials at the $\lambda_{(e,g)}$'s.

\begin{construction}\label{cons2}
For each data file of size $B$, the node $(e,g)$ stores the $\bar{d}$ symbols:
$f_0(\lambda_{(e,g)}),f_1(\lambda_{(e,g)}),...,f_{\bar{d}-1}(\lambda_{(e,g)})\;.$ In other words, the code $C$ can be expressed as
{\setlength\abovedisplayskip{6pt}
	\setlength\belowdisplayskip{6pt}
\begin{equation}\label{cons}C=M\Lambda\end{equation}}
where $M$ is the $\bar{d}\times k$ matrix derived from the data file, $\Lambda=(\lambda_{(e,g)}^j)_{j,(e,g)}$ with row index $j\in [0,k-1]$ and column index $(e,g)\in[0,\bar{n}-1]\times[0,u-1]$, and $C$ is a $\bar{d}\times n$ code matrix such that each node stores a column of $C$.
\end{construction}

Since all the polynomials defined in (\ref{polys}) have degree at most $k-1$, by Lagrange interpolation any $k$ nodes can recover these polynomials whose coefficients form the original data file. Namely, any $k$ nodes can reconstruct the data file. In the following we prove the code $C$ constructed in (\ref{cons}) satisfies the optimal repair property. In short, the node repair of $C$ relies on the node repair of the product-matrix MBR code \cite{Kumar2011} and the local repair property within each rack.

Firstly we illustrate the local repair property.
For each rack $e\in[0,\bar{n}-1]$ and $i\in[0,\bar{d}-1]$, define a polynomial
$h^{(e)}_i(x)=\sum_{j=0}^{u-1}h^{(e)}_{i,j}x^j$
where
\begin{equation}\label{poly}
h_{i,j}^{(e)}=\begin{cases}
\sum_{t=0}^{\bar{k}}m_{i,tu+j}\cdot\xi^{etu}\ \ \ \ \ \ \ \mathrm{if} \ 0\leq j<u_0\\
\sum_{t=0}^{\bar{k}-1}m_{i,tu+j}\cdot\xi^{etu}\ \ \ \ \ \ \ \mathrm{if} \ u_0\leq j<u-1\\
\sum_{t=0}^{\bar{d}-1}m_{i,tu+u-1}\cdot\xi^{etu}\ \ \ \ \mathrm{if} \ j=u-1
\end{cases}\;.
\end{equation}

\begin{lemma}\label{lemma} For all
$e\in[0,\bar{n}-1]$ and $i\in[0,\bar{d}-1]$, it holds $f_i(\lambda_{(e,g)})=h^{(e)}_i(\lambda_{(e,g)})$ for $g\in[0,u-1]$.
\end{lemma}
\begin{proof}
For $j\!\in\! [0,k\!-\!1]$, write $j\!=\!tu\!+\!\nu$ where $t\!=\!\lfloor\frac{j}{u}\rfloor$. Next we rearrange the terms of $f_i(\lambda_{(e,g)})$ according to the values of $\nu$, i.e.,
\begin{align*}
f_i(\lambda_{(e,g)})=&\sum_{\nu=0}^{u_0-1}\sum_{t=0}^{\bar{k}}m_{i,tu+\nu}\lambda_{(e,g)}^{tu+\nu}+\sum_{\nu=u_0}^{u-2}\sum_{t=0}^{\bar{k}-1}\\
&m_{i,tu+\nu}\lambda_{(e,g)}^{tu+\nu}\!+\!\sum_{t=0}^{\bar{d}-1}m_{i,tu+u-1}\lambda_{(e,g)}^{tu+u-1}.
\end{align*}
Note from (\ref{ms2}) it has $m_{i,tu+u-1}=0$ for $t\in[\bar{d},\bar{k}-1]$.
Then because $\lambda_{(e,g)}=\xi^e\eta^g$ and $\eta$ has multiplicative order $u$, it has $\lambda_{(e,g)}^{tu}=\xi^{etu}$. Combining with the definition in (\ref{poly}), it is easy to verify
$f_i(\lambda_{(e,g)})=\sum_{\nu=0}^{u-1}h_{i,\nu}^{(e)}\lambda_{(e,g)}^\nu=h_{i}^{(e)}(\lambda_{(e,g)})$.	
\end{proof}

\begin{remark}\label{Re2}
Lemma \ref{lemma} shows that for each rack $e\in[0,\bar{n}-1]$, when restricted to the $u$ nodes within rack $e$, the punctured code $C_e$ is actually defined by $\bar{d}$ polynomials of degree at most $u-1$, i.e., $h^{(e)}_i(x)$, $i\in[0,\bar{d}-1]$. If the leading coefficients of the $h^{(e)}_i(x)$'s are already known, then any node erasure can be recovered from the remaining $u-1$ nodes in the same rack. We call this as the local repair property. The idea of reducing the polynomial degree at each local group to insure the local repair property has been used in constructing LRCs \cite{LRCFamily2014}. Here we further extend the idea to constructing linear array codes combining with the product-matrix  MBR codes \cite{Kumar2011} for storing the leading coefficients in the cross-rack level.
\end{remark}

\begin{lemma}\label{lemma2}
For $e\!\in\![0,\bar{n}\!-\!1]$, let ${\bm h}_e$ be the vector composed of the leading coefficients of $h^{(e)}_i(x)$ for $i\in[0,\bar{d}-1]$, i.e.,  $${\bm h}_e=(h_{0,u-1}^{(e)},h_{1,u-1}^{(e)},...,h_{\bar{d}-1,u-1}^{(e)})^\tau\in F^{\bar{d}}\;.$$
Then $$({\bm h}_{0},{\bm h}_{1},...,{\bm h}_{\bar{n}-1})=S\begin{pmatrix}1&1&\cdots&1\\1&\xi^{u}&\cdots&\xi^{(\bar{n}-1)u}\\\vdots&\vdots&\vdots&\vdots\\1&(\xi^{u})^{\bar{d}-1}&\cdots&(\xi^{(\bar{n}-1)u})^{\bar{d}-1}
\end{pmatrix}\;,$$
where $S$ is the $\bar{d}\times\bar{d}$ symmetric matrix in (\ref{ms2}).
Therefore, $({\bm h}_{0},{\bm h}_{1},...,{\bm h}_{\bar{n}-1})$ is actually a codeword of an $(\bar{n},\bar{d},\bar{d})$ MBR code.
\end{lemma}

\begin{proof}
The proof is directly from the expression in (\ref{poly}) and the product-matrix construction of MBR codes in \cite{Kumar2011}.
\end{proof}

\begin{theorem}
Given the code $C$ constructed in (\ref{cons}), any node erasure can be recovered from all the remaining $u-1$ nodes in the same rack, as well as $\bar{d}$ helper racks each transferring $\beta=1$ symbol.

In other words, any column of $C$ indexed by $(e^*,g^*)$ can be recovered from the $u-1$ columns indexed by $\{(e^*,g):0\leq g\leq u-1,g\neq g^*\}$ and $\bar{d}$ symbols each of which is a linear combination of the columns of the punctured code $C_e$ for $\bar{d}$ helper racks $e\neq e^*$.
\end{theorem}

\begin{proof}
For any node erasure $(e^*,g^*)$ and any $\bar{d}$ helper racks $e_1,...,e_{\bar{d}}\!\in\![0,\bar{n}\!-\!1]\!-\!\{e^*\}$, the MBR code proved in Lemma \ref{lemma2} implies that ${\bm h}_{e^*}$ can be recovered from $\bar{d}$ symbols ${\bm \lambda}_{e^*}^\tau{\bm h}_{e_i}, 1\!\leq\! i\!\leq\! \bar{d}$, where $${\bm \lambda}_{e^*}^\tau=(1,\xi^{e^*u},(\xi^{e^*u})^2,...,(\xi^{e^*u})^{\bar{d}-1})\;.$$
Moreover, by Lemma \ref{lemma} ${\bm h}_{e_i}$ is a linear combination of the columns of $C_{e_i}$ for $1\leq i\leq \bar{d}$, and the erased column $C_{(e^*,g^*)}$ is a linear combination of ${\bm h}_{e^*}$ and the other $u-1$ columns of $C_{e^*}$. Thus the theorem follows.
\end{proof}

\section{Comparisons and Conclusions}
In this paper, we establish a tradeoff between the storage overhead and cross-rack repair bandwidth for the rack-aware storage system for all parameters. In particular, we characterize the parameters of the MSRR codes and MBRR codes in the case $\bar{d}<\bar{k}$ and also present explicit constructions for the two kinds of codes. Next we make some comparisons between the MSRR/MBRR codes at different parameters to give some advice in selecting parameters in practice. As an illustration, Table \ref{t0} lists the pairs of storage overhead and cross-rack repair bandwidth overhead, i.e.,  $(\frac{n\alpha}{B},\frac{\bar{d}\beta}{\alpha})$, at different values of $\bar{n}$ and $\bar{d}$ for fixed $u$ and $n-k$.
\begin{table}[ht]
	\renewcommand\arraystretch{1.5}
	\setlength{\abovecaptionskip}{0.cm}
	\setlength{\belowcaptionskip}{-0.cm}
	\caption{
\scriptsize Fix $u\!=\!5$ and $n\!-\!k\!=\!6$. The table lists the pairs of $(\frac{n\alpha}{B},\frac{\bar{d}\beta}{\alpha})$ at different values of $\bar{n}$ and $\bar{d}$. The codes can be built over ${\rm GF}(2^8)$ by Construction \ref{cons1} or \ref{cons2}.}
	\begin{center}
		\resizebox{0.9\columnwidth}{!}{
			\begin{tabular}{|c|c|c|c|c|c|}
				\hline
				\multirow{2}{*}{}
				&$\bar{d}=0$&\multicolumn{2}{c|}{$\bar{d}=4$}&\multicolumn{2}{c|}{$\bar{d}=8$}\\
				\cline{2-6}  & MSRR & MSRR & MBRR & MSRR & MBRR \\
				\hline $\bar{n}=10$ & $(1.389,0)$ & $(1.25,4)$ & $(1.299,1)$ & $(1.136,8)$ & $(1.235,1)$ \\
				\hline $\bar{n}=20$ & $(1.316,0)$ & $(1.25,4)$ & $(1.274,1)$ & $(1.190,8)$ & $(1.242,1)$ \\
				\hline $\bar{n}=30$ & $(1.293,0)$ & $(1.25,4)$ & $(1.266,1)$ & $(1.210,8)$ & $(1.245,1)$ \\
				\hline
		\end{tabular}}
	\end{center}
\label{t0}
\end{table}

\vspace{-4pt}
\begin{enumerate}
  \item ${\bar{d}}\geq \bar{k}$ vs. ${\bar{d}}< \bar{k}$. Of these two choices, we always prefer the latter. Because in a system with low storage overhead, $\bar{k}$ is usually near to $\bar{n}$, for example, $\bar{k}=\bar{n}-2,\bar{n}-3$, etc. Thus ${\bar{d}}\geq \bar{k}$ leads to a rather high repair degree which causes much rack-connectivity in the repair process and even makes the repair cannot work when multiple failures happen. Besides, the MSRR codes with ${\bar{d}}\geq \bar{k}$ are generally realized by array codes with exponential sub-packetication, while our MSRR codes with ${\bar{d}}< \bar{k}$ are scalar linear codes. The main advantage of choosing ${\bar{d}}\geq \bar{k}$ is the lower storage overhead. However, the codes with ${\bar{d}}< \bar{k}$ can also have satisfiable storage overhead (see Table \ref{t0}).

  \item MSRR code vs. MBRR code in the case $0\!<\!{\bar{d}}\!<\!\bar{k}$. From Table \ref{t0} one can see the MBRR code has prominent advantage over the MSRR code in saving the cross-rack repair bandwidth, although there is a small sacrifice in the storage overhead. Thus we prefer MBRR codes in most cases.

  \item $0\!<\!\bar{d}\!<\!\bar{k}$ vs. ${\bar{d}}\!=\!0$. As stated in Remark \ref{re1} our MSRR codes in the case $\bar{d}=0$ actually induce optimal LRCs. Although $\bar{d}\!>\!0$ introduces cross-rack communication in the repair process, it reduces the storage overhead to some extend. As listed in Table \ref{t0}, when $\bar{n}=10$ from $\bar{d}=0$ to $\bar{d}=8$ the storage overhead of MSRR codes is reduced by 18.2\%, and that of MBRR codes is reduced by 11.1\%. Moreover, the cross-rack repair bandwidth overhead of MBRR codes is always $1$. Therefore, choosing MSRR codes with $\bar{d}=0$ or MBRR codes with small $\bar{d}$ depends on the system is more sensitive to the cross-rack connectivity or storage overhead.
\end{enumerate}

\end{document}